\documentclass[10pt]{iopart}

\newcommand{\A}{_{\mathrm{a}}}
\newcommand{\B}{_{\mathrm{b}}}

\newcommand{\Z}{_{\mathrm{z}}}
\newcommand{\Ll}{_{\mathrm{L}}}
\newcommand{\Cc}{_{\mathrm{c}}}

\usepackage{iopams}  
  \usepackage{cite}
  \usepackage{graphicx}

\begin{document}

\title[Zitterbewegung in a fluctuating optical lattice]{Zitterbewegung with spin-orbit coupled ultracold atoms
in a fluctuating optical lattice}

\author{V.~Yu.~Argonov, D.~V.~Makarov}

\address{
Pacific Oceanological Institute of the Russian Academy of Sciences,\\
43 Baltiiskaya st., 690041 Vladivostok, Russia, URL: http://dynalab.poi.dvo.ru}  

\begin{abstract}

Dynamics of non-interacting ultracold atoms with artificial spin-orbit coupling is considered.
Spin-orbit coupling is created using two moving optical lattices with orthogonal polarizations.
Our main goal is to study influence of lattice noise on Rabi oscillations. 
Special attention is paid to the phenomenon of the Zitterbewegung being
trembling motion caused by Rabi transitions between states with different velocities. 
Phase and amplitude fluctuations of lattices are modelled by means 
of the two-dimensional stochastic Ornstein-Uhlenbeck process, 
also known as harmonic noise.
In the the noiseless case the problem is solved analytically
in terms of the momentum representation.
It is shown that lattice noise significantly extends duration of the Zitterbewegung as compared
to the noiseless case. This effect originates from noise-induced 
decoherence of Rabi oscillations.

\end{abstract}

%
%
%
%
\ioptwocol
\section{Introduction}
\label{Intro}

In recent years, enormous progress had been achieved in experimental realization of artificial gauge fields
with ultracold atoms \cite{Lin_Spielman09,Aidelsburger-PRL11}. It gives rise to the new era in quantum simulation 
of solid-state phenomena, such as
the quantum Hall effect \cite{Sorensen-PRL05,Kolovsky-JPB13,JPB13}, 
spin-orbit coupling \cite{Lin_Spielman11,Hamner-PRL15}, Majorana fermions \cite{Sato-PRL09}, to name a few.
This breakthrough is mainly conditioned by high degree of atom state controllability, allowing to reduce
undesirable concomitant factors which lead to fast decoherence of quantum states in solid-state experiments.
However, we have to account for mechanismes of decoherence which are relevant for ultracold atoms, like
optical lattice amplitude noise \cite{Pichler-PRA12,Pichler-PRA13} or spontaneous emission \cite{PRA08,EPL08,Schachenmayer-PRA14}. 

In the this work we consider dynamics of ultracold atoms with artificial spin-orbit coupling
in the presence of optical lattice fluctuations. Attention is focused on influence of noise on Rabi inter-level
oscillations. It is well-known that Rabi coupling for spin-orbit coupled states leads to onset of 
the Zitterbewegung (ZB) oscillations. 
The term ZB means specific trembling motion originally predicted by Schr\"odinger
for relativistic Dirac electrons \cite{Schrod-ZB}.
Basically, ZB occurs as a consequence of coupling between states with different velocities.
The phenomenon of the ZB with ultracold atoms was considered, for example, in Refs. \cite{Qu-PRA13,LeBlanc-ZB}.
We are interested in influence of lattice amplitude fluctuations on ZB oscillations.
We consider the experimental setup where spin-orbit coupling is created by means of the Raman dressing scheme \cite{Lin_Spielman11}.
Lattice fluctuations are introduced using the harmonic noise \cite{HN,Anischenko}.
In the present work we restrict ourselves by the case of non-interacting atoms.

The paper is organized as follows. In Section \ref{Model} we introduce the model under consideration.
Section \ref{Momentum} represents the analytical solution for Rabi oscillations in the absence of noise.
Noise influence on Rabi oscillations is studied in Section \ref{Rabi}.
Effect of the Zitterbewegung is considered in Section \ref{ZB}.
In Section \ref{Discuss} we summarize the results obtained and outline possible ways for further research.


\section{Model}
\label{Model}

Consider gas of ultracold non-interacting ${}^{87}$Rb atoms with mass $M$ and momentum $p$
moving in an external magnetic field and an optical field of two Raman lasers
with orthogonal linear polarizations. 
Frequency difference between the two lasers is $\omega\Ll$ and the $x$-projection of wave vector difference is $k\Ll$.
Magnetic field results in Zeeman splitting of the triplet state $F=1$.
Laser field induces transitions between three hyperfine states $m_F=-1$, $m_F=0$, $m_F=1$. 
Setting laser frequency difference $\omega\Ll$ to be close to the frequency splitting $\omega\Z$
between $m_F=-1$ and $m_F=0$ states, we can neglect
the level $m_F=1$ due to the quadratic Zeeman effect. Therefore we have an effective two-level ($|m_F=-1\rangle$, $|m_F=0\rangle$) system with
the Hamiltonian \cite{Martone-PRA12}
\begin{equation}
\hat H = \frac{\hat p^2}{2M} + \frac{\hbar\Omega}{2}[\hat\sigma_1U_1(x,t) +
\hat\sigma_2U_2(x,t)] - \frac{\hbar\omega_z}{2}\hat\sigma_3,
\label{Ham}
\end{equation}
where 
$\hat\sigma_{1,2,3}$ are the Pauli matrices and $\Omega$ is the Rabi frequency (fixed by the intensity of the lasers). 
The Hamiltonian (\ref{Ham}) is characterized by equal contributions of Rashba and Dresselhaus couplings.
In the absence of lattice fluctuations,
the terms $U_1$ and $U_2$ read
\begin{equation}
\eqalign{
 U_1(x,t) &= \cos(2k\Ll x-\omega\Ll t),\\
 U_2(x,t) &= \sin(2k\Ll x-\omega\Ll t),}
 \label{Udet}
\end{equation}
Lattice fluctuations can be incorporated into the Hamiltonian 
by means of the replacement \cite{PLA,EPJB14,PhysScr15}
\begin{equation}
\eqalign{
 U_1(x,t)= f(t)\cos(2k\Ll x) - sf(t+\tau)\sin(2k\Ll x),\\
 U_2(x,t)= f(t)\sin(2k\Ll x) + sf(t+\tau)\cos(2k\Ll x),}
 \end{equation}
where $f(t)$ is the two-dimensional Ornstein-Uhlenbeck process, also called
harmonic noise. Harmonic noise is described by
the coupled stochastic differential equations
\begin{equation}
 \dot f=y,\quad
 \dot y=-\Gamma y-\omega\Ll ^2f + \sqrt{2\Gamma}\xi(t),
 \label{ou2d}
\end{equation}
where $\Gamma$ is a positive constant, and
$\xi(t)$ is Gaussian white noise.
It should be emphasized that 
terms $f(t)$ and $f(t+\tau)$ have to correspond to the same realization of noise $\xi$
in order to provide wavelike form of $U_1$ and $U_2$ \cite{PLA}.

Power spectrum of harmonic noise has an unique peak at frequency 
\begin{equation}
 \omega_{\mathrm{p}}=\sqrt{\omega\Ll ^2-\frac{\Gamma^2}{2}}.
\label{w_p}
\end{equation}
Width of the peak is given by
\begin{equation}
\Delta\omega = \sqrt{\omega_{\mathrm{p}}^2+\Gamma\omega'}-
\sqrt{\omega_{\mathrm{p}}^2-\Gamma\omega'},
 \label{width}
\end{equation}
where 
\begin{displaymath}
\omega'=\sqrt{\omega\Ll ^2-\frac{\Gamma^2}{4}}.  
\end{displaymath}
It is important to note that elimination of the hyperfine level $m_F=1$ requires 
\begin{equation}
 |\omega\Ll-\omega\Z|\gg \Delta\omega.
 \label{valid}
\end{equation}
If $\Gamma \ll \omega\Ll$, then
\begin{equation}
 \Delta\omega \approx \Gamma.
\end{equation}
In the present work we consider values of $\Gamma$ in the range $10^{-3}\omega\Ll\div 10^{-1}\omega\Ll$.
If we set $\omega\Ll=4.81$ MHz, as in \cite{Lin_Spielman11}, the corresponding laser bandwidth should be $10^3\div 10^5$ Hz.

In the deterministic limit $\Gamma\to 0$ we have
\begin{displaymath}
 f(t)\to \cos(\omega\Ll t+\phi_0).
\end{displaymath}
Choosing initial conditions in (\ref{ou2d}) as $f(0)=1$, $y(0)=0$, we specify $\phi_0=0$.
Then setting 
\begin{equation}
 \tau=\frac{\pi}{2\omega\Ll },
 \label{tau}
\end{equation}
one reproduces (\ref{Udet}) if $\Gamma=0$. 
Hence, it turns out that $U_1(x,t)$ and $U_2(x,t)$
for $\Gamma>0$ behave as fluctuating plane waves. 
Let's denote
\begin{equation}
f(t) + if(t+\tau) = W(t)e^{-i\varphi(t)}.
\label{Wdef}
\end{equation}
In the deterministic case $\Gamma=0$ we have
$ \varphi = \omega\Ll t$.

After the replacement (\ref{Wdef}), the Schr\"odinger equation 
can be represented as a pair of coupled equations
\begin{equation}
\eqalign{
 i\hbar\frac{\partial \Psi\A}{\partial t} =& \left(\frac{\hat p^2}{2M} - \frac{\hbar\omega\Z}{2}\right)\Psi\A+
 \frac{\hbar\Omega W}{2}e^{i(\varphi-2k\Ll x)}\Psi\B,\\
 i\hbar\frac{\partial \Psi\B}{\partial t} =& \left(\frac{\hat p^2}{2M} + \frac{\hbar\omega\Z}{2}\right)\Psi\B+
 \frac{\hbar\Omega W}{2}e^{-i(\varphi-2k\Ll x)}\Psi\A,}
\label{sys} 
\end{equation}
where $\Psi\A$ and $\Psi\B$ are wave functions of the $|m_F=-1\rangle$ and $|m_F=0\rangle$ states, respectively.

Assume that dynamics is restricted by spatial domain of length $L$ being integer multiple $N$ of the lattice period,
i.~e.
\begin{equation}
 L = \frac{\pi N}{k\Ll},\quad N\gg 1.
\end{equation}
Imposing periodic boundary conditions at the ends of the domain,
we can define discrete set of momentum eigenfunctions
\begin{equation}
 \psi_n = \frac{1}{\sqrt{L}}\exp(-ip_nx/\hbar),
 \end{equation}
where $p_n = n\hbar k_0$, $n$ is integer, $k_0 = 2\pi/L$.
Expanding $\Psi\A$ and $\Psi\B$ over momentum eigenfuctions
\begin{equation}
 \Psi\A = \sum_n \alpha_n(t)\psi_n,\quad
 \Psi\B = \sum_n \beta_n(t)\psi_n,
 \label{pexp}
\end{equation}
we can significantly simplify the problem by reducing
(\ref{sys}) to pairs of coupled ODE
\begin{equation}
\eqalign{
 i\frac{d\alpha_n}{dt} = \frac{\Omega W}{2}\exp\left[i\left(\varphi-\nu_n t\right)\right]\beta_{n+l},\\
 i\frac{d\beta_{n+l}}{dt} = \frac{\Omega W}{2}\exp\left[-i\left(\varphi-\nu_n t\right)\right]\alpha_n,}
\label{ab}
 \end{equation}
where 
\begin{equation}
\eqalign{
 \nu_n = \frac{E\B(n+l) - E\A(n)}{\hbar},\\
 E_{\mathrm{a,b}}(n) = \frac{p_n^2}{2M} \mp \frac{\hbar\omega\Z}{2},\quad
 l = \frac{2k\Ll}{k_0}.}
\label{defin}
\end{equation}
For simplicity, we use scaling of variables corresponding to $M=1$. 

\section{Noiseless case -- analytical solution}
\label{Momentum}

In the purely deterministic case $\Gamma=0$  solution of the equations (\ref{ab}) is given by
\begin{equation}
\eqalign{
\alpha_n = \sqrt{\rho_n}\left(c_{n1}e^{i\Omega_n^+ t}+c_{n2}e^{i\Omega_n^- t}\right)e^{i\gamma_n}, \\
\beta_{n+l} = \sqrt{\rho_n}\left(c_{n3}e^{-i\Omega_n^- t}+c_{n4}e^{-i\Omega_n^+ t}\right)e^{i\gamma_n},
}
\label{sol}
 \end{equation}
where $\gamma_n$ is phase of a corresponding term in the expansion (\ref{pexp}), $c_{jn}$ are real-valued coefficients,
\begin{equation}
\Omega_n^{\pm}\equiv\frac{1}{2}\left(\chi_n\pm \tilde{\Omega}_n\right),\quad
\tilde{\Omega}_n \equiv \sqrt{\chi_n^2+\Omega^2},
  \label{Ompm}
\end{equation}
\begin{equation}
 \chi_n \equiv\omega\Ll -\nu_n.
 \label{chi}
\end{equation}
%
Constants $c_{nj}$ obey the following relation:
\begin{equation}
  c_{n1}^2+c_{n2}^2+c_{n3}^2+c_{n4}^2=1.
  \label{norm}
\end{equation}
Amplitudes $\alpha_n$ and $\beta_{n+l}$ satisfy the conservation law
\begin{equation}
 |\alpha_n|^2 + |\beta_{n+l}|^2 = \rho_n.
\end{equation}
Substituting (\ref{sol}) into (\ref{ab}), we find the relations
\begin{equation}
c_{n3} = -\frac{2\Omega_n^+}{\Omega}c_{n1},\quad
c_{n4} = -\frac{2\Omega_n^-}{\Omega}c_{n2}.
\label{c34}
\end{equation}
We use the initial condition of a form
\begin{equation}
|\alpha_n(t=0)| = \sqrt{\rho_n},\quad
|\beta_n(t=0)| = 0
 \label{init}
\end{equation}
for all $n$. Then we find
\begin{equation}
\eqalign{
 c_{n1} = -\frac{\Omega_n^-}{\tilde{\Omega}_n},\quad
 c_{n2} = \frac{\Omega_n^{+}}{\tilde{\Omega}_n},\\
 c_{n3} = -\frac{\Omega}{2\tilde{\Omega_n}} = -c_{n4}.}
 \label{cn}
\end{equation}
The case of 
\begin{equation}
 \chi_n=0
 \label{res}
\end{equation}
corresponds to onset of resonance in Eqs.~(\ref{ab}), when $c_{n1}=c_{n2}=-c_{n3}=c_{n4}=1/2$,
therefore, $\alpha_n$ and $\beta_{n+l}$ oscillate with frequency $\Omega/2$.
Accordingly, we can regard the parameter $\chi_n$ as detuning of resonance between
coupled momentum states. Increasing of $\chi_n$ results in growing of $|c_{n2}|$, while 
$|c_{n1}|$, $|c_{n3}|$ and $|c_{n4}|$ decrease.

Link between $\chi_n$ and corresponding momentum value can be easily derived from (\ref{defin}) and (\ref{chi}):
\begin{equation}
\chi_n = \omega\Ll - \omega\Z - 2k\Ll p_n - 2\hbar k\Ll^2.
 \label{chi-p}
\end{equation}
Any localized quantum state has finite width in the momentum space
and therefore consists of many momentum states.
In particular, we consider the initial state with Gaussian-like distribution
\begin{equation}
\rho_n = \frac{e^{-(p_n-p\Cc)^2/2\sigma_p^2}}
{\sum_n e^{-(p_n-p\Cc)^2/2\sigma_p^2}}.
\label{rho_n} 
\end{equation}
Owing to the linear dependence of $\chi_n$ on $p_n$,
only one of the momentum states may satisfy 
the condition (\ref{res}), other ones correspond to off-resonant
transitions with nonzero values of $\chi_n$.

\section{Phase space representation of Rabi oscillations}
\label{Rabi}

\begin{figure}[!tb]
\begin{center}
\includegraphics[width=0.48\textwidth]{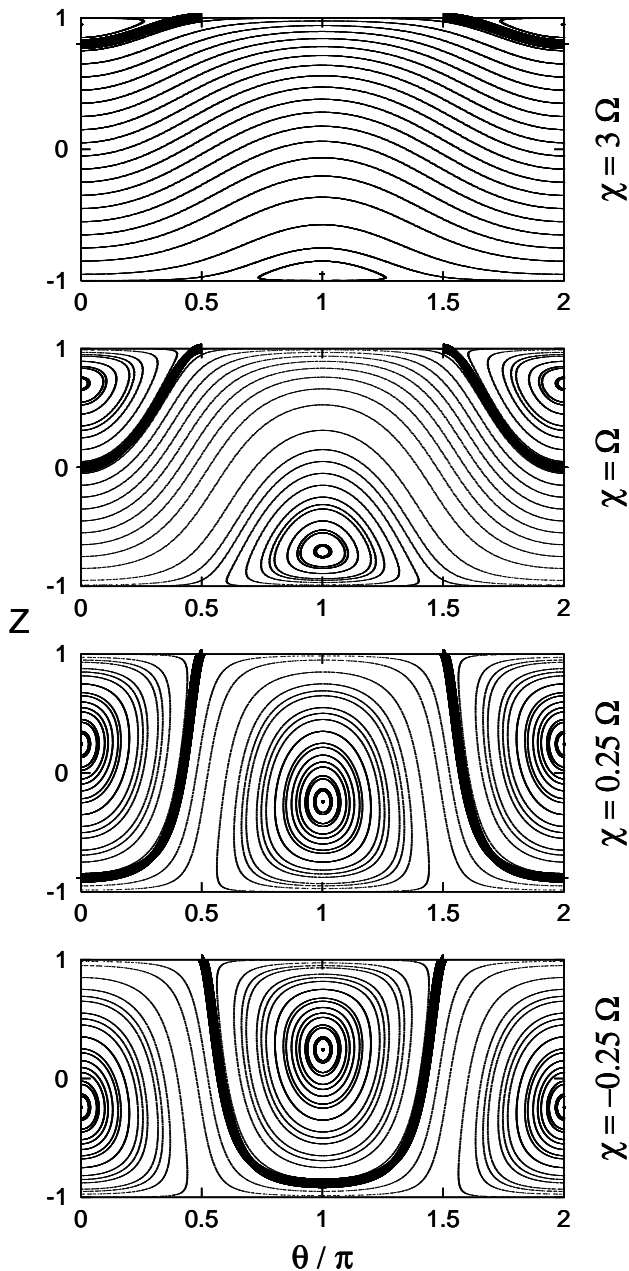}
\caption{
Phase portraits for the Hamiltonian (\ref{H}) in the absence of noise.
Values of the parameter $\chi$ are indicated to the right of each panel.
Bold line corresponds to the solution (\ref{sol}) with coefficients determined by (\ref{cn}).
}
\label{Fig-php}
\end{center}
\end{figure}

Rabi oscillations between resonantly coupled momentum components 
of $|m_F=-1\rangle$ and $|m_F=0\rangle$ states can be readily described  by means of 
normalized population imbalance for a $n$-th momentum state
\begin{equation}
 Z(n,t) = \frac{|\alpha_n(t)|^2 - |\beta_{n+l}(t)|^2}{\rho_n}
 \label{zn}
\end{equation}
%
After some simple algebra with the usage of (\ref{sol}) and (\ref{cn}), 
we find the solution for $Z(n,t)$ in the absence of noise
\begin{equation}
Z = \frac{1}{\tilde\Omega_n^2}(\chi_n^2 + \Omega^2\cos\tilde\Omega t)
\label{zn0}
\end{equation}
Further we omit the subscript $n$. From Eq.~(\ref{zn0}) it is clear that quantity $\tilde\Omega$ stands
for effective Rabi frequency taking into account momentum-dependent detuning from the resonance (\ref{res}).

In the presence of noise, dynamics of $Z$ is governed by the equation that comes from (\ref{ab})
and looks as
\begin{equation}
 \frac{dZ}{dt} = 
 -\Omega W(t)\sqrt{1-Z^2}\sin\theta,
 \label{dzdt}
 \end{equation}
where
\begin{equation}
\theta = \arg{\alpha} - \arg{\beta} - \varphi + \nu t. 
\end{equation}
Eq.~(\ref{dzdt}) is complemented by equation
\begin{equation}
 \frac{d\theta}{dt}=\Omega W(t)\frac{Z}{\sqrt{1-Z^2}}\cos\theta - \chi - \eta_{\chi}(t).
 \label{dth}
\end{equation}
where $\eta_{\chi}(t)=\dot{\varphi}(t)-\omega\Ll$ is a fluctuating part of 
the frequency difference between the Raman lasers.
Equations (\ref{dzdt}) and (\ref{dth}) can be rewritten in the Hamiltonian form
\begin{equation}
 \frac{dZ}{dt} = -\frac{\partial H}{\partial\theta},\quad
 \frac{d\theta}{dt} = \frac{\partial H}{\partial Z},
 \label{Hsys}
\end{equation}
with the Hamiltonian
\begin{equation}
 H = -\Omega (1+\eta_{\mathrm{w}})\sqrt{1-Z^2}\cos\theta - (\chi + \eta_{\chi}) Z,
 \label{H}
\end{equation}
where we used the replacement $W(t) = 1+\eta_{\mathrm{w}}(t)$.
Form of the Hamiltonian (\ref{H}) is generic for problems studying population dynamics.
For example, a similar expression is used for describing population dynamics
in external \cite{Raghavan} and internal Bose-Josephson junctions \cite{Oberthaler1,JRLR14}.

Stationary points of the Hamiltonian (\ref{H}) 
in the absence of noise $\eta_{\mathrm{w}}=\eta_{\chi}=0$ can be found by solving the equations
\begin{equation}
 \frac{dZ}{dt}=\frac{d\theta}{dt}=0.
\end{equation}
This gives 
\begin{equation}
 \theta_{\mathrm{st}} = \pi (2k+1),\quad
 Z_{\mathrm{st}} = -\frac{\chi}{\tilde\Omega},
 \label{st1}
\end{equation}
and
\begin{equation}
 \theta_{\mathrm{st}} = 2\pi k,\quad
 Z_{\mathrm{st}} = \frac{\chi}{\tilde\Omega},\quad
 k\in \mathbb{Z}.
 \label{st2}
\end{equation}
Eqs. (\ref{st1}) and (\ref{st2}) determine equillibrium points for Rabi oscillations with the dominance
of the $|m_F=0\rangle$ or $|m_F=-1\rangle$ states, depending on the sign of $\chi$.
Figure \ref{Fig-php} illustrates phase portraits for Rabi oscillations with various values of $\chi$.
Equillibrium points are placed within the island-like phase space regions where phase $\theta$ is trapped. 
Despite the solution (\ref{sol}) goes by the trapping regions, the corresponding trajectory
lies predominantly in the upper half-plane, inferring prevailence of the state $|m_F=-1\rangle$.
The only exception is the case of exact resonance $\chi=0$ (not shown in the Figure).
Dominance of the state $|m_F=-1\rangle$
is especially apparent in the case of $\chi=3\Omega$, when the trajectory only slightly deviates from the upper bound $Z=1$.
Indeed, far from resonance (\ref{res}) right-hand side terms in (\ref{ab}) rapidly oscillate
with time and can be averaged out. Therefore, Rabi oscillations are suppressed.
A similar phenomenon, referred to as the coherent population trapping, was earlier observed
in \cite{Prants96,Prants-OS97,JETP09,Konkov_Prants09} for two- and three-level atoms moving in a standing-wave laser field.

As noise is turned on, a phase space trajectory corresponding to Rabi oscillations 
is able to diffuse in phase space. 
So, it can  drift to a region with different population dynamics, or even enter a trapping region.
This leads to intermittency in Rabi oscillations.
Phase space diffusion tends to mix the regimes where one of the state dominates, therefore, one may expect that
time- and ensemble-averaged population imbalance 
\begin{equation}
 \bar Z = \frac{1}{N_{\mathrm{r}}T}\sum\limits_{j=1}^{N_{\mathrm{r}}}\int\limits_{t=0}^T \bar Z^{(j)}(t)\,dt,
 \label{averz}
\end{equation}
should be closer to zero in the presence of noise than in the noiseless case.
In Eq.~(\ref{averz}),
superscript $j$ labels the realizations of harmonic noise, and $N_{\mathrm{r}}$ is total number of realizations.
We calculate $\bar Z$ for initial conditions (\ref{init}).
In the absence of noise, it can be easily found from (\ref{zn}) that 
\begin{equation}
 \bar Z = \frac{\chi^2}{\chi^2 + \Omega^2}.
 \label{bZ}
\end{equation}
\begin{figure}[!tb]
\begin{center}
\includegraphics[width=0.48\textwidth]{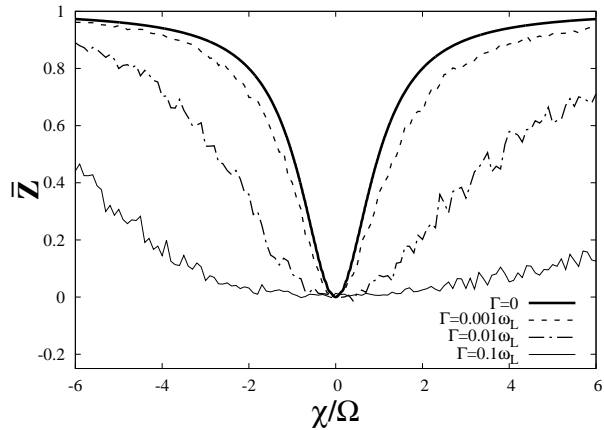}\\
\caption{
Time-averaged population imbalance between the $|m_F=-1\rangle$ and $|m_F=0\rangle$ states
for the $n$-th momentum state
as function of the parameter $\chi_n$ defined by (\ref{chi}).
Corresponding values of the harmonic noise parameter $\Gamma$ 
are indicated in the corner of the Figure.
Parameters values: $\omega\Ll=10\Omega$, length of the interval for time averaging is $T=100\pi/\Omega$.
The curves for nonzero $\Gamma$ correspond to averaging over 120 realizations
of harmonic noise.
}
\label{Fig-dz}
\end{center}
\end{figure}
The corresponding curve
is shown in Fig.~\ref{Fig-dz} by the bold line.
As $\chi$ approaches to resonance, $\bar Z$ decreases from 1 to 0 at $\chi=0$.
Thus, we can regard function $\bar Z(\chi)$ as resonance curve describing the process
of the photon absorption by a two-level system.
Numerical simulation in the presence of noise shows
that intermittency reduces values of the time-averaged population imbalance as compared
to the noiseless case.
This leads to remarkable broadening of the resonance absorption peak,
as $\Gamma$ grows. This is especially pronounced in the case of $\Gamma = 0.1\Omega$.

\section{Zitterbewegung}
\label{ZB}

Population imbalance is related to a momentum expectation value 
$ p(t) =  \langle\Psi\A|\hat p|\Psi\A\rangle + \langle\Psi\B|\hat p|\Psi\B\rangle$
by means of the formula
\begin{equation}
 p(t) = p(t=0) + \hbar k\Ll\left[1-\sum_n \rho_n Z(n,t)
  \right],
  \label{pt}
\end{equation}
where it is taken into account that $p_{n+l}-p_n = 2\hbar k\Ll$.
As we use scaling corresponding to $M=1$,
$p$ is equal to velocity of center-of-mass motion.
As it follows from (\ref{pt}), Rabi inter-level transtions result in oscillations of $p$, 
hence, there occurs trembling motion referred to as the Zitterbewegung (or shortly ZB).
Form of ZB oscillations depends on 
the initial momentum distribution (\ref{rho_n}).
Firstly, it depends on the width of the distribution, as interference of terms with different $n$
in (\ref{pt}) should damp ZB.
In numerical simulation, we set $\sigma_p=0.01\hbar k\Ll$. Owing to the Heisenberg uncertainty relation, it corresponds
to $\sigma_x=\hbar/(2\sigma_p)\simeq 8\lambda\Ll$, where $\lambda\Ll$ is laser wavelength.
The center of the momentum distribution $p\Cc$ is taken of 0.
Secondly, oscillations of $p$ depend on the detuning of the initial state 
from the resonance (\ref{res}).
Let's introduce mean detuning as
\begin{equation}
\chi\Cc = \omega\Ll - \omega\Z - 2k\Ll p\Cc - 2\hbar k\Ll^2.
\label{chi_c} 
\end{equation}
\begin{figure}[!htb]
\begin{center}
\includegraphics[width=0.4\textwidth]{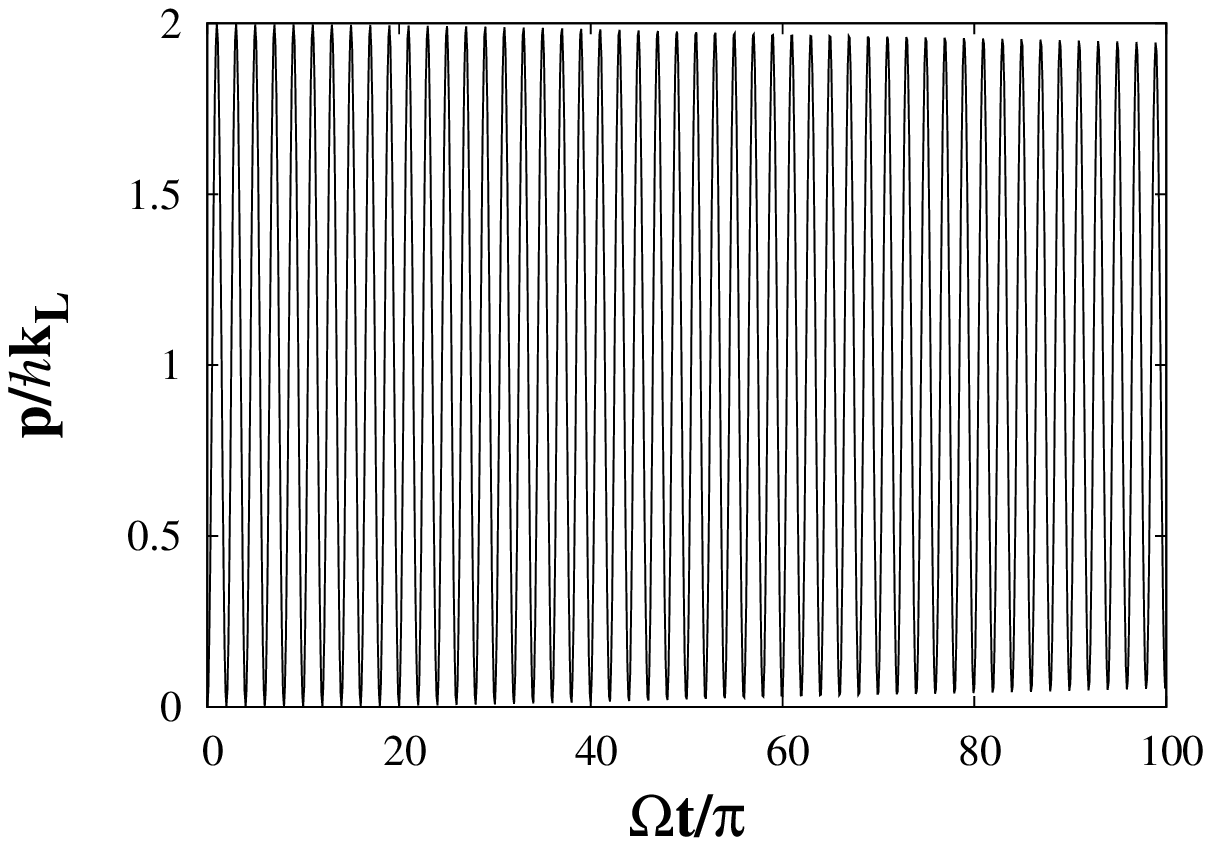}
\includegraphics[width=0.4\textwidth]{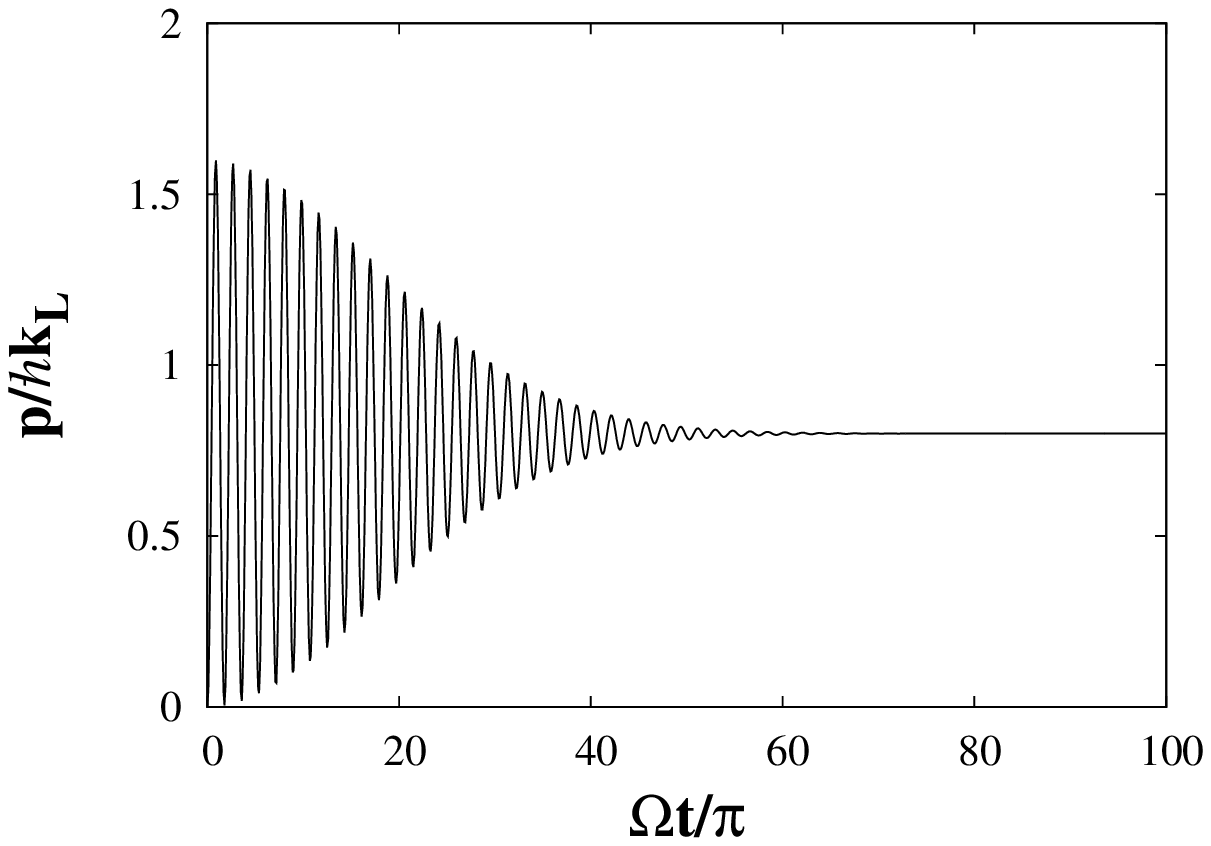}
\caption{Quantum expectation value of momentum as function of time for $\chi\Cc=0$ (upper panel) 
and $\chi\Cc=0.5\Omega$ (lower panel).
}
\label{Fig-zb}
\end{center}
\end{figure}
Figure \ref{Fig-zb} represents ZB oscillations in the absence of noise for $\chi\Cc=0$ (upper panel) and $\chi\Cc=0.5\Omega$ (lower panel).
Notably, ZB exposes much faster damping in the case of $\chi\Cc=0.5\Omega$
than in the case of $\chi\Cc=0$ corresponding to the strongest influence of resonance.

\begin{figure}[!htb]
\begin{center}
\includegraphics[width=0.48\textwidth]{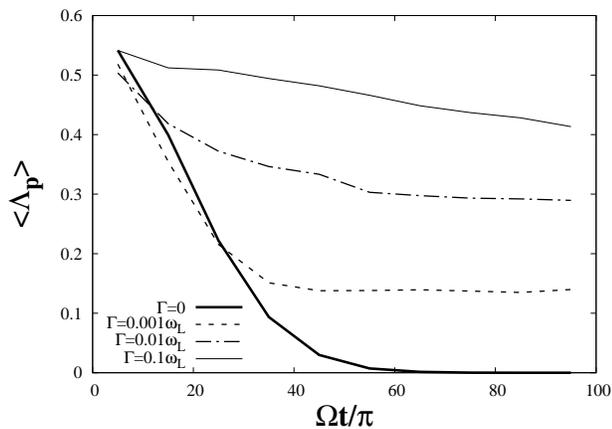}
\caption{Standard deviation of momentum within a temporal interval of length $\Delta t=10\pi/\Omega$.
Values of other parameters:
$\omega\Ll=10\Omega$, $\chi\Cc=0.5\Omega$.
}
\label{Fig-rmsp}
\end{center}
\end{figure}

Now let's consider effect of noise on ZB oscillations.
In the presence of noise,
strength of the ZB phenomenon can be quantified by amplitude of momentum oscillations.
Therefore, it is reasonable to consider 
standard deviation of $p$ within a temporal interval of length $\Delta t$
\begin{equation}
\eqalign{
\Lambda_p(t) = \frac{1}{\hbar k\Ll}\sqrt{\overline{p^2} - \overline{p}^2},\\
\overline{p^k}(t) \equiv \frac{1}{\Delta t}\int\limits_{t-\Delta t/2}^{t+\Delta t/2}p^k(t')\,dt',\quad k=1,2.
}
 \label{Lambda}
\end{equation}
The length of the temporal interval $\Delta t$ has to be much larger than 
 characteristic period of inter-level transitions $T\approx 2\pi/\Omega$. In numerical simulation we set $\Delta t=10\pi/\Omega$.
Data for $\Lambda_p$ is averaged over 100 realizations of harmonic noise.

\begin{figure}[!htb]
\begin{center}
\includegraphics[width=0.48\textwidth]{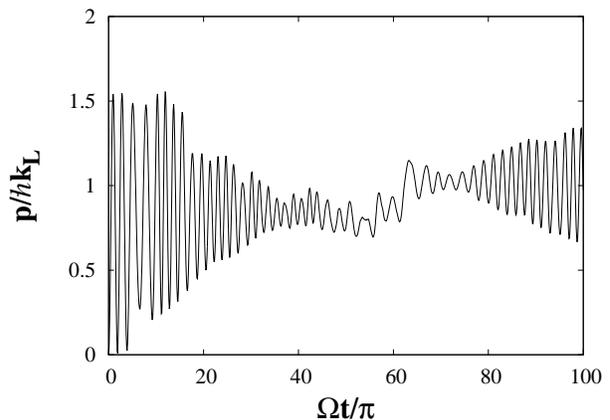}
\caption{Momentum expectation value as function of time for a single realization of harmonic noise with 
$\Gamma=0.001\omega\Ll$. The case of $\chi\Cc=0.5\Omega$.
}
\label{Fig-zbnoise}
\end{center}
\end{figure}

Evidently noise should add stochasticity into Rabi oscillations.
One may expect that noise-induced phase fluctuations of Rabi oscillations can reduce phase coherence 
that is responsible for the ZB damping in the noiseless case, in analogy with noise-induced destruction of the Anderson localization 
\cite{EPJB14,Derrico,DeFalco_Tamascelli-2013}.
This expectation is fully confirmed by results of numerical simulation presented in Fig.~\ref{Fig-rmsp}.
The case of $\chi\Cc=0.5\Omega$ is considered.
After decreasing within the initial stage, ensemble-averaged value of $\Lambda_p$ achieves a plateau for all non-zero
values of $\Gamma$, indicating persistence of ZB.
Although increasing of $\Gamma$ leads to higher values of $\Lambda_p$, 
effect of noise-induced ZB persistence is apparent 
even in the case of very weak noise $\Gamma=10^{-3}\omega\Ll$.
To understand the origin of the plateau, let's consider 
a typical realization of $p(t)$ for $\Gamma=10^{-3}\omega\Ll$.
Such an example is presented in Fig.~\ref{Fig-zbnoise}.
One can see that ZB oscillations firstly demonstrate damping, as in the noiseless case, but then there arises
a burst with growing amplitude, as phase coherence of Rabi oscillations for different momentum eigenstates pairs
is destructed.

\section{Summary}
\label{Discuss}

In the present work we consider dynamics of non-interacting ultracold atoms with artificial spin-orbit
coupling imposed. Spin-orbit coupling is realized by means of the Raman dressing.
Atoms move in a superposition of two moving optical lattices which experience
random amplitude and phase fluctuations. 
Our main concern is to study influence of fluctuations on inter-level Rabi oscillations, with 
particular emphasis on the phenomenon of the Zitterbewegung (ZB).  
We find an analytical solution for the noiseless case. It is shown that 
dynamics of Rabi oscillations can be represented by an autonomous Hamiltonian system with one degree of freedom.
As detuning from resonance (\ref{res}) is significantly deviated from zero,
there can occur  the effect of the population trapping. 
In the presence of noise the corresponding Hamiltonian system has 3/2 degrees of freedom. 
Noise results in the onset of intermittency in Rabi oscillations.
The main result of the paper is the noise-induced deceleration of ZB damping that is observed even if noise is weak.
We link this effect with destruction of coherence in Rabi oscillations for different momentum components.

We model the fluctuations of a laser field by means of the harmonic noise.
It implies that they are caused by some uncontrollable factors reducing laser coherence.
However, the same role can be played by properly chosen broadband modulation of the laser wave.
We expect that the effect of noise-induced persistence for ZB oscillations can be observed
with broadband deterministic signals as well. 
Also, it is very interesting how this effect will change in the presence
of interaction between atoms, for example, within a mean-field picture. We intend to address these issues
in the forthcoming works.


\section*{Aknowledgments}

This work was partially supported by the Russian Foundation of Basic Research
within the projects 15-02-08774-a and 15-02-00463-a.

\section*{References}


\providecommand{\newblock}{}

\end{document}